\documentclass{elsart}
\usepackage{graphicx}
\usepackage{dcolumn}
\usepackage{bm}

\begin{document}

\begin{frontmatter}

\title{The breathing-mode giant monopole resonance and 
the surface compressibility in the relativistic mean-field theory}

\author{M.M. Sharma}
\address{Physics Department, Kuwait University, Kuwait 13060}

\ead{sharma@kuc01.kuniv.edu.kw}
\date{\today}

\begin{abstract}
The breathing-mode isoscalar giant monopole resonance (GMR) is investigated 
using the generator coordinate method within the relativistic mean-field (RMF) 
theory. Employing the Lagrangian models of the 
nonlinear-$\sigma$ model (NL$\sigma$), the scalar-vector 
interaction model (SVI) and the $\sigma$-$\omega$ coupling  
model (SIGO), we show that each Lagrangian 
model exhibits a distinctly different GMR response. Consequently,  
Lagrangian models yield a different value of the GMR energy for a given 
value of the nuclear matter incompressibility $K_\infty$. It is shown that 
this effect arises largely from a different value of the surface 
incompressibility $K_{surf}$ inherent to each Lagrangian model, 
thus giving rise to the ratio $K_{surf}/K_\infty$  which depends 
upon the Lagrangian model used. This is attributed to a difference in the 
density dependence of the meson masses and hence to the 
density dependence of the nuclear interaction amongst 
various Lagrangian models. The sensitivity of the GMR energy to the 
Lagrangian model used and thus emergence of a multitude of GMR energies 
for a given value of $K_\infty$ renders the method 
of extracting $K_\infty$ on the basis of interpolation amongst forces as 
inappropriate. As a remedy, the need to 'calibrate' the density dependence 
of the nuclear interaction in the RMF theory is proposed.

\end{abstract}


\end{frontmatter}

\maketitle

\section{Introduction}
\label{Intro}

The compressibility of nuclear matter has been a matter of exploration
and discussion for a long time. The incompressibility or the compression 
modulus is a fundamental property of the nuclear matter. It constitutes 
a cardinal point on the equation of state (EOS) of the nuclear matter. 
Defined as the second derivative of the EOS of the nuclear matter 
at the saturation density, it is also important for astrophysical 
phenomena such as supernova explosion and structure of neutron 
stars \cite{bethe.90,Hillebrandt.85,Baron.87}.

The breathing-mode isoscalar giant monopole resonance (GMR) is a 
principal source of information on the (in)compressibility of nuclei 
and nuclear matter \cite{Blaizot.80,Treiner.81,Harakeh.01}. 
Nuclei undergo radial density oscillations about the equilibrium 
point in this mode. It is a small-amplitude
collective motion wherein the nucleus participates as a whole.
The isoscalar GMR is well-established experimentally and has been 
studied in a large number of nuclei from heavy to light mass regions
in various laboratories around the world 
\cite{Buenerd.84,Youngblood.81,Lu.86,Sharma.88}. Recent measurements 
\cite{Youngblood.99,Li.07} have brought in improved precision in 
providing better information on the breathing-mode GMR. 

Experimental determination of the incompressibility of nuclear matter
has entailed using a liquid-drop model type expansion of the
incompressibility $K_A$ of a nucleus in terms of a bulk and finite-size 
components \cite{Buenerd.84,Sharma.88}. Experimental data on the 
breathing-mode isoscalar GMR on a large number of nuclei was 
used \cite{Buenerd.84}. Attempts were made to extract the nuclear matter 
incompressibility $K_\infty$ using the data on Sn and Sm isotopes 
obtained with considerable precision \cite{Sharma.88}. However, the 
correlation between the Coulomb term and $K_\infty$ hinders  
extraction of the incompressibility of the nuclear matter 
\cite{Pearson.91}.

With the difficulty of extracting $K_\infty$ directly using the
experimental data, theoretical approach based upon interpolation
amongst Skyrme and Gogny type interactions as proposed by Blaizot 
\cite{Blaizot.80,Blaizot.95} has been  extensively employed. 
It has required reproducing experimental breathing-mode 
GMR energies of nuclei using an interaction with an appropriate
value of $K_\infty$ in  self-consistent Hartree-Fock and random-phase
approximation (RPA) calculations \cite{Blaizot.95}. 
In this scheme, the GMR energy of $^{208}$Pb has been shown to be 
reproduced with forces with $K_\infty \sim 220$ MeV. 
On the other hand, the RPA calculations
with force(s) which reproduce the GMR energy of $^{208}$Pb give much
larger value for $^{90}$Zr, another key nucleus, as compared to the 
experimental data \cite{Blaizot.95}.
The GMR energy of $^{90}$Zr \cite{Youngblood.04}  
has since been overestimated by $\sim 1-2$ MeV in nonrelativistic 
\cite{Agrawal.05} and relativistic \cite{Pieka.04} RPA approach.
In the same vain, the GMR energies of Sn and Sm isotopes could not 
be described within the RPA sum-rule approach \cite{Gleissl.90, Sharma.89}
using the force SkM* with $K_\infty \sim 216$ MeV, which described 
$^{208}$Pb successfully though. It has been pointed out recently that
the GMR energies of Sn isotopes are overestimated in a self-consistent
RPA approach by 1-2 MeV by the forces which reproduce the GMR energy of
$^{208}$Pb appropriately \cite{Sagawa.07,Pieka.07}. 

Using the RMF approach and relativistic RPA, a similar behaviour has 
been observed \cite{Ma.00}. The success of the RPA approach and 
extraction of $K_\infty$ using the interpolation method thus hinges 
very strongly on the nucleus $^{208}$Pb which masks
attendant problems with other nuclei. 

With the advent of the relativistic mean-field (RMF) theory 
\cite{SW.86,Rein.89,Ser.92,Ring.96}, properties of nuclei
and nuclear matter can be studied within this framework
which has become successful in describing ground-state 
properties of nuclei along and far from the stability line 
\cite{GRT.90,SNR.93,Lala.97}. The inherent advantage 
of the RMF theory lies in the Dirac-Lorentz structure of nucleons, 
which allows a built-in spin-orbit interaction. Consequently, in 
properties related to shell structure, especially that of
the anomalous behaviour of the isotope shifts of Pb nuclei, 
the RMF theory exhibits special advantages over the  
nonrelativistic Skyrme theory \cite{SLR.93}.  

The breathing-mode isoscalar GMR has been studied within the RMF
theory earlier \cite{Sto.94}. Again, the theoretical or 
microscopic method of extracting $K_\infty$ entails 
an interpolation amongst various forces with different 
values of $K_\infty$ with a view to being able to reproduce GMR 
energies of nuclei such as $^{208}$Pb, Sn isotopes
and $^{90}$Zr among others within a microscopic approach 
\cite{Vretenar.03}.  The Lagrangian model with nonlinear scalar 
self-couplings of $\sigma$ meson has mostly been used
with this approach. It has been pointed out in a schematic
approach that forces with $K_\infty \sim 250-270$ in the RMF model 
with the nonlinear scalar self-couplings (of the form 
$\sigma^3 + \sigma^4$) would be compatible with the 
experimental GMR energy of $^{208}$Pb \cite{Vretenar.03}.
However, it has been shown recently that the GMR data on Sn 
isotopes and $^{90}$Zr were overestimated by such a force \cite{Pieka.07}.
Relativistic RPA calculations have also been performed with 
Lagrangians with density-dependent meson couplings \cite{Niksic.02},
which focus only on the key nucleus of $^{208}$Pb.

The interpolation scheme has been the cornerstone of the theoretical
method to extract the incompressibility of nuclear matter 
microscopically. Using the various Lagrangian models developed,
it will be shown in this work that this approach runs into problems 
due to a sensitivity shown by GMR energies to the Lagrangian model 
employed. A difference in the density dependence of the meson masses 
amongst the various Lagrangian models and an ensuing difference in
the response of the surface of nuclei to compression is ascribed for 
this behaviour.

The subject matter of the paper is organized as follows:
Definitions of the basic ingradients of the GMR and the incompressibility 
of the nuclear matter are given in Section \ref{GMR}. In Section
\ref{formalism}, the formalism of the RMF theory is provided. We 
discuss the Lagrangian models used in this work in Section \ref{RMF}. 
Details of the generator coordinate method employed 
to explore and calculate the GMR energies of nuclei are
discussed in Section \ref{GCM}. It will be shown in Section \ref{results}
that the GMR energies exhibit a sensitivity to the Lagrangian model 
considered. We shall show that the surface incompressibility 
plays an important role for the GMR energies in the RMF theory. 
The last section summarizes the main results and
an outlook for future investigations is presented. 

\section{The breathing-mode giant monopole resonance and the compressibility
of nuclear matter}

\label{GMR}

The isoscalar (T=0) giant monopole resonance (GMR) has been 
well established in the last few decades \cite{Harakeh.01}.
It is a breathing-mode of a nucleus
wherein the nucleus undergoes a density oscillation about its
equilibrium value. The breathing mode GMR energy or its frequency 
is related to the incompressibility $K_A$ of a nucleus with mass 
number $A$ by
\begin{equation}
E_0 = \hbar \sqrt\frac{K_A}{m\langle r^2\rangle},
\label{E0}
\end{equation}
where $m$ is the nucleon mass and $\langle r^2 \rangle$ is 
the mean-square mass radius of the nucleus in its ground state. 

Analogous to the expansion of the total energy of a nucleus
in terms of a liquid drop formula, one can write the expansion 
of the incompressibility of the nucleus in terms of contributions from 
the volume, surface, asymmetry, Coulomb and curvature terms, 
respectively, as \cite{Blaizot.80}:
\begin{equation}
\begin{array}{rl}
K_A = & \sum K_i c_i \\
    = & K_\infty + K_{surf}A^{-1/3} + K_{asym} \left( \frac{N-Z}{A} \right)^2 
+ K_{Coul}\frac{Z^2} {A^{4/3}} + K_{curv} A^{-2/3},
\label{KA}
\end{array}
\end{equation}
where $N$ and $Z$ are the neutron and proton numbers, respectively. 
The $K_i$'s are 'incompressibilities' (coefficients) representing various 
terms and $c_i$'s are the terms of the expansion in various powers of $A$.
The volume term $K_\infty$ is defined as the curvature of the EOS of the 
infinite nuclear matter:
\begin{equation}
\label{kinf}
K_\infty = 9\rho_0^2 \frac{d^2 E/A}{d\rho^2}\vert_{\rho_0},
\end{equation} 
where the second-derivative of the EOS, i.e., of $\epsilon(\rho)$ is 
evaluated at the saturation density $\rho_0$. Thus, $K_\infty$ 
represents the most significant (reference) point of the 
high-density behaviour of the infinite nuclear matter. 

The volume and surface terms provide the most significant contributions to
$K_A$, whereas the asymmetry and the Coulomb terms are much smaller in 
comparison. The curvature term is even smaller than the latter two.
The volume term being the largest contributor to $K_A$ influences the 
breathing-mode GMR energies of nuclei directly.

The Coulomb term $K_{Coul}$ has been derived as \cite{Blaizot.80} 
\begin{equation}
\label{Coul}
K_{Coul} = \frac{3}{5}\frac{e^2}{r_0}\left(1 - 
\frac{27\rho_0^2}{K_\infty}{e'''}\vert_{\rho_0}\right),
\end{equation}
where $e''' = d^3(E/A)/d\rho^3$ is the third-derivative 
(skewness) of the EOS. As readily seen 
from Eq. (\ref{Coul}), $K_{Coul}$ is correlated to $K_\infty$. 
In Section \ref{results} we shall see  that the magnitude of the coefficient
$K_{Coul}$ is about two orders of magnitude smaller than the major
contributors such as $K_\infty$ and $K_{surf}$. Due to this 
correlation superimposed on to the meagre quantity of the 
Coulomb term, it is difficult to extract $K_\infty$ from 
breathing-mode energies \cite{Pearson.91}. In the present work, we shall 
circumvent this correlation and attempt to disentangle the various 
terms.

\section{The Relativistic Mean-Field Theory - Formalism}
\label{formalism}

The starting point of the RMF theory is the 
basic Lagrangian (the linear Walecka model) that describes 
nucleons as Dirac spinors interacting with the meson fields \cite{SW.86}:
\begin{eqnarray}
\label{lagran}
{\cal L}_0&=& \bar\psi \left( \rlap{/}p - g_\omega\rlap{/}\omega -
g_\rho\rlap{/}\vec\rho\vec\tau - \frac{1}{2}e(1 - \tau_3)\rlap{\,/}A -
g_\sigma\sigma - m\right)\psi\nonumber\\
&&+\frac{1}{2}\partial_\mu\sigma\partial^\mu\sigma
   -\frac{1}{2}m^2_\sigma \sigma^2
   -\frac{1}{4}\Omega_{\mu\nu}\Omega^{\mu\nu}+ \frac{1}{2}
m^2_\omega\omega_\mu\omega^\mu\\ &&
-\frac{1}{4}\vec R_{\mu\nu}\vec R^{\mu\nu}+
\frac{1}{2} m^2_\rho\vec\rho_\mu\vec\rho^\mu -\frac{1}{4}F_{\mu\nu}F^{\mu\nu},
\nonumber
\end{eqnarray}
where $m$ is the bare nucleon mass and $\psi$ is its Dirac spinor. 
Nucleons interact with the $\sigma$, $\omega$, and $\rho$ mesons. 
Here, $g_\sigma$, $g_\omega$, and $g_\rho$ are the respective coupling 
constants of the interaction. The photonic field is produced by the 
electromagnetic vector $A^\mu$. 

The field tensors of the vector mesons and of the electromagnetic
field are given by:
\begin{equation}
\begin {array}{rl}
\label{field.T}
\Omega^{\mu\nu} =& \partial^{\mu}\omega^{\nu}-\partial^{\nu}\omega^{\mu}\\
{\bf R}^{\mu\nu} =& \partial^{\mu}
                  \mbox{\boldmath $\rho$}^{\nu}
                  -\partial^{\nu}
                  \mbox{\boldmath $\rho$}^{\mu}\\
F^{\mu\nu} =& \partial^{\mu}{\bf A}^{\nu}-\partial^{\nu}{\bf A}^{\mu}.\\
\end{array}
\end{equation}
The variational principle gives rise to the Dirac equation:
\begin{equation}
\label{Dirac}
\left( -i{\bf {\alpha}}.\nabla + V({\bf r}) + \beta m*  \right)
~\psi_{i} = ~\epsilon_{i} \psi_{i},
\end{equation}
where $V({\bf r})$ represents the $vector$ potential:
\begin{equation}
\label{vpot}
V({\bf r}) = g_{\omega} \omega_{0}({\bf r}) + g_{\rho}\tau_{3} {\bf {\rho}}
_{0}({\bf r}) + e{1-\tau_{3} \over 2} {A}_{0}({\bf r}),
\end{equation}
and $S({\bf r})$ is the $scalar$ potential
\begin{equation}
\label{spot}
S({\bf r}) = g_{\sigma} \sigma({\bf r}).
\end{equation}
The effective mass is defined by the scalar potential as
\begin{equation}
m^{\ast}({\bf r}) = m + S({\bf r}).
\end{equation}
The corresponding Klein-Gordon equations can be written as:
\begin{equation}
\begin{array}{rl}
\label{KG1}
( -\Delta + m_{\sigma}^{*2} ) \sigma = & -g_\sigma\bar\psi\psi \\
( -\Delta + m_{\omega}^{*2} ) \omega_\nu = & g_\omega\bar\psi\gamma_\nu\psi \\
( -\Delta + m_{\rho}^{*2} ) \vec\rho_\nu = 
           & g_{\rho} \bar\psi\gamma_\nu\vec\tau\psi \\
 -\Delta A_\nu = & \frac{1}{2}e\bar\psi(1 + \tau_3)\gamma_\nu\psi.
\end{array}
\end{equation}
For the linear Lagrangian model,
\begin{equation}
\label{eff}
m_\sigma^* =  m_\sigma; ~~~m_\omega^* =  m_\omega; ~~~m_\rho^*   =  m_\rho. 
\end{equation}
Thus, the meson masses do not exhibit a density dependence in the
linear model. For an even-even nucleus with time-reversal symmetry, 
the spatial components of the vector fields 
\mbox{\boldmath $\omega$}, \mbox{\boldmath $\rho_3$} and
\mbox{\boldmath A} vanish. The Klein-Gordon equations for the meson 
fields are then time-independent inhomogeneous equations with the 
nucleon densities as sources:
\begin{equation}
\begin{array}{rl}
\label{KG2}
( -\Delta + m_{\sigma}^{2} )\sigma({\bf r})
  =& -g_{\sigma}\rho_{s}({\bf r}) \\
  (-\Delta + m_{\omega}^{2} ) \omega_{0}({\bf r})
  =& g_{\omega}\rho_{v}({\bf r}) \\
    ( -\Delta + m_{\rho}^{2} )\rho_{0}({\bf r})
  =& g_{\rho} \rho_{3}({\bf r})\\
  -\Delta A_{0}({\bf r}) = &e\rho_{c}({\bf r}).
\end{array}
\end{equation}
where the source terms $\rho_s$, $\rho_v$, $\rho_3$ and $\rho_p$ to the 
Klein-Gordon equations (Eq. \ref{KG2}) are the scalar, vector, 
isovector and charge densities, respectively, as defined by nucleon spinors:
\begin{equation}
\begin{array}{ll}
\label{dens}
\rho_{s} =& \sum\limits_{i=1}^{A} \bar\psi_{i}~\psi_{i}\\
\rho_{v} =& \sum\limits_{i=1}^{A} \psi^{+}_{i}~\psi_{i}\\
\rho_{3} =& \sum\limits_{p=1}^{Z}\psi^{+}_{p}~\psi_{p}~-~
\sum\limits_{n=1}^{N} \psi^{+}_{n}~\psi_{n}\\
\rho_{c} =& \sum\limits_{p=1}^{Z} \psi^{+}_{p}~\psi_{p}.
\end{array}
\end{equation}
Here, the sums are taken over the valence nucleons only. The stationary state 
solutions $\psi_i$ are obtained from the coupled system of Dirac 
(Eq. \ref{Dirac}) and Klein-Gordon equations (Eq. \ref{KG2}) 
self-consistently. 

The total ground-state energy of a spherical nucleus can be expressed
as a functional of the baryon spinors $\{\psi_i\}$
\begin{equation}
\label{deq}
E_{RMF} [\psi] \equiv \langle \Phi \vert \hat{H} \vert \Phi \rangle,
\end{equation}
obtained using the Hamiltonian density

\begin{eqnarray}
\label{hamilton}
{\cal H}_{RMF}(r) &=& \tau(r) + m\rho_s(r) \nonumber\\
        && + \frac{1}{2} g_\sigma\rho_s(r)\sigma(r) 
         +\frac{1}{2}g_\omega\rho_v(r)\omega^0(r) \nonumber \\
&& +\frac{1}{2}g_\rho\rho_3(r)\rho^0(r) +\frac{1}{2}e\rho_p(r)A^0(r)\nonumber\\
&& +\frac{1}{2}((\nabla\sigma(r))^2 + m_\sigma^2\sigma^2(r)\nonumber\\
&& -\frac{1}{2}((\nabla\omega^0(r))^2 + m_\omega^2(\omega^0(r))^2\nonumber\\
&& -\frac{1}{2}((\nabla\rho^0(r))^2 + m_\rho^2(\rho^0(r))^2\nonumber\\
&& -\frac{1}{2}((\nabla A^0(r))^2 
\end{eqnarray}
which is given in terms of the source densities (Eq. \ref{dens}). The
kinetic energy density is given by
\begin{equation}
\tau(r) \equiv \sum\limits_{i=1}^{A} \psi_i^\dagger(r)\{-i{\bf\alpha \nabla}
\} \psi_i(r).
\end{equation}
Taking the variation of Eq. (\ref{deq}) with respect to $\psi_i^\dagger$,
the stationary Dirac equation (Eq. \ref{Dirac}) with energy 
eigenvalues $\epsilon_i$ is obtained:
\begin{equation}
\label{dd}
\hat{h}_D\psi_i(r) = \epsilon_i\psi_i(r), 
\end{equation}
where
\begin{eqnarray}
\hat{h}_D = -i{\bf\alpha \nabla} + \beta m^\ast 
           +g_\omega\omega^0(r) + g_\rho\tau_3\rho^0(r) 
             +e\frac{(1-\tau_3)}{2}A^0(r).
\end{eqnarray}
Solving the Dirac equation (Eq. \ref{dd}) self-consistently, 
the ground-state $\Phi_0$ of the nucleus  is written  as a Slater 
determinant  of single-particle spinors $\psi_i$ (i = 1,2,...,A). 

The linear Walecka model has been successful in attaining 
saturation of nuclear matter as a delicate 
balance between large fields due to $\sigma$ and 
$\omega$ mesons. However, a proper description of the properties of 
finite nuclei was not possible until nonlinear self-couplings of
the $\sigma$ meson were introduced \cite{BB.77}. 

\section{The RMF Lagrangian models}
\label{RMF}

In this work, we have considered three successful RMF Lagrangian models 
in order to analyze the breathing-mode isoscalar GMR. In the following, we
provide a brief description of the formulation of the various Lagrangian
models considered.

\subsection{The nonlinear-$\sigma$ model}

The nonlinear sigma (NL$\sigma$) model is the standard Lagrangian
model that is used most commonly for calculation of the ground-state properties
of nuclei. An important ingredient of the NL$\sigma$ model is the 
assumption of nonlinear scalar self-couplings of the form \cite{BB.77}:
\begin{equation}
\label{nlsig}
U_{NL} = \frac{1}{3}g_2\sigma^3 + \frac{1}{4}g_3\sigma^4. 
\end{equation}
The parameters $g_2$ and $g_3$ are the nonlinear couplings of the 
$\sigma$-meson in the conventional $\sigma^3$ + $\sigma^4$ model. 
The effective Lagrangian for the NL$\sigma$ model becomes
\begin{equation}
\label{laneff}
{\cal L}_{eff} = {\cal L}_0 - U_{NL}.
\end{equation}
The Klein-Gordon equations (Eq. \ref{KG1}) produce a density
dependence of the meson masses as given by
\begin{equation}
\begin{array}{rl}
\label{effm1}
m_{\sigma}^{*2} = & m_{\sigma}^2 + g_2\sigma 
                   + g_3{\sigma}^2  \\
m_\omega^*  = & m_\omega \\
m_\rho^*    = & m_\rho.
\end{array}
\end{equation}
Only the $\sigma$-meson mass exhibits an implicit density dependence
in the NL$\sigma$ model. 

The scalar self-couplings have proved to be important for an appropriate 
description of nuclear surface and have thus
become indispensable. However, the negative quartic coupling in the model 
NL$\sigma$ has been a source of instability in nuclear matter at higher
densities \cite{Rein.89,Furnstahl.87,Sulaksono.07}. It has been shown 
recently \cite{Sharma.08} that a scalar-vector interaction (SVI) model 
that comprises a combination of couplings of $\sigma$ and $\omega$ 
mesons can dispense with the $\sigma^3 + \sigma^4$ terms in the RMF 
Lagrangian. The SVI model is able to provide an improved
description of the ground-state binding energies and charge radii of 
nuclei. We shall discuss the basic features of the SVI model in this 
section below.

The NL$\sigma$ model is well-established and has shown to be a successful
model for calculating ground-state properties of nuclei. The earliest
parameter sets obtained were NL1 ($K_\infty = 211$ MeV) \cite{Rein.89}
and NL2 ($K_\infty = 399$ MeV) \cite{Lee.86}. 
However, due to the large asymmetry energy $J$,
these sets were not appropriate for nuclei away from the stability line.
In remedying the problem of a large neutron skin \cite{Sharma.92},
the force NL-SH \cite{SNR.93} was developed as one of the first 
successful parameter sets in the RMF theory, which also described nuclei 
away from the stability line. Due to a relatively larger 
value of $K_\infty$ of NL-SH, the force NL3 has been obtained \cite{Lala.97}
with a view to getting a description of the breathing-mode 
GMR in a physically acceptable region.
\noindent
\begin{table}[h*]
\begin{center}
\caption{The nuclear matter properties of various parameter sets in the 
NL$\sigma$ model.}
\bigskip
\begin{tabular}{l l c c c c c }
\hline
& Sets     & $K_\infty$ (MeV) & $E/A$ (MeV) & $m^*$ 
& $\rho_0$ (fm$^{-3}$) & $J$ (MeV) \\    
\hline 
& NL1               & 211   & -16.42 & 0.573   & 0.152  & 43.5 \\  
& NL3               & 272   & -16.25 & 0.595   & 0.148  & 37.4 \\
& NL-SH             & 355   & -16.32 & 0.597   & 0.146  & 36.1 \\ 
& NL2               & 399   & -17.02 & 0.667   & 0.146  & 45.1 \\
\hline
\end{tabular}
\end{center}
\vspace{0.5cm}
\end{table}

In this work, we consider the forces NL1, NL3, NL-SH and NL2, 
whose order reflects an increasing value of $K_\infty$. Nuclear matter
properties of these parameter sets are provided in Table 1.
It may be noted that even though the forces NL1 and NL2 are quite 
different in some of the properties such as incompressibility 
and asymmetry energy, these are able to reproduce ground-state 
properties of nuclei along the stability line quite well. 
Here, we have included NL2 in order to extend the systematic behaviour
of the NL$\sigma$ model into the region of high incompressibility.

\subsection{The $\sigma$-$\omega$ coupling model - SIGO} 

The $\sigma$-$\omega$ coupling model (SIGO) has recently been
introduced by Haidari and Sharma \cite{Haidari.08} with a
view to bring about an improvement in the ground-state properties of
nuclei. A coupling between $\sigma$ and $\omega$ mesons of
the form 
\begin{equation}
\label{sigo}
U_{\sigma\omega} = \frac{1}{2}g_{\sigma\omega} \sigma^2\omega^2
\end{equation}
was introduced in addition to the usual NL$\sigma$ scalar
potential of the form $\sigma^3$ + $\sigma^4$.
The effective Lagrangian for the model SIGO becomes
\begin{equation}
\label{laneff2}
{\cal L}_{eff} = {\cal L}_0 - U_{NL} + U_{\sigma\omega}.
\end{equation}
The meson mass terms in the corresponding Klein-Gordon equations 
(Eq. \ref{KG1}) are then given by:
\begin{equation}
\begin{array}{rl}
\label{effm2}
m_{\sigma}^{*2} = & m_{\sigma}^2 + g_2\sigma 
                   + g_3{\sigma}^2 - g_{\sigma\omega}{\omega_0}^2 \\
m_{\omega}^{*2} = & m_{\omega}^2 + g_{\sigma\omega}\sigma^2 \\
m_\rho^*        = & m_\rho.
\end{array}
\end{equation}
Thus, in the SIGO model, both the $\sigma$ and $\omega$ meson masses 
exhibit an implicit density dependence.

It has been shown \cite{Haidari.08} that the parameter set SIG-OM 
obtained in the model SIGO is able to provide an excellent description 
of the ground-state properties such
as binding energies and charge radii of nuclei along the stability
line as well as far away from it.  Especially, a significant 
improvement in the binding energy of nuclei at the magic numbers
has been achieved. Consequently, an excellent description of the 
total binding energy of Sn isotopes all over the range of the shell
from $^{100}$Sn ($N=50$) to $^{132}$Sn ($N=82$) was 
obtained. Charge radii of nuclei, especially of Pb isotopes, 
which are over estimated significantly by NL3, were well
reproduced. The ensuing EOS of the nuclear matter with 
SIG-OM has been shown to be much softer than that given by 
the parameter sets of the NL$\sigma$ model \cite{Haidari.08}.

In order to investigate the behaviour of the model SIGO for
the breathing-mode GMR, we have constructed a few parameter 
sets by extending the range of $K_\infty$ below and above 
that of the set SIG-OM ($K_\infty = 265$ MeV).
The parameter sets SIGO-a, SIGO-b, SIGO-c and SIGO-d have been
obtained with $K_\infty$ ranging between 241 - 283 MeV.
The nuclear matter properties of the parameter sets
of the model SIGO are given in Table 2. It may be noted that
the sets SIGO-a, SIGO-b, SIGO-c and SIGO-d corresponding to
different values of $K_\infty$ are able to reproduce the
binding energy and charge radii of key nuclei from $^{16}$O
to $^{208}$Pb well. Qualitatively, the set SIG-OM  
is considered as the best amongst all the sets of the SIGO model 
provided in Table 2. 

\noindent
\begin{table}[h*]
\begin{center}
\caption{The nuclear matter properties of the $\sigma$-$\omega$ 
coupling (SIGO) sets.}
\bigskip
\begin{tabular}{l l c c c c c }
\hline
& Sets     & $K_\infty$ (MeV) & $E/A$ (MeV) & $m^*$ 
& $\rho_0$ (fm$^{-3}$) & $J$ (MeV) \\    
\hline 
& SIGO-a            & 241.1 & -15.90 & 0.621   & 0.149  & 33.0 \\  
& SIGO-b            & 248.4 & -16.00 & 0.622   & 0.150  & 33.6 \\
& SIG-OM            & 265.2 & -16.30 & 0.622   & 0.149  & 37.0 \\
& SIGO-c            & 272.8 & -16.33 & 0.623   & 0.149  & 37.7 \\ 
& SIGO-d            & 282.5 & -16.34 & 0.620   & 0.148  & 37.5 \\
\hline
\end{tabular}
\end{center}
\end{table}

\subsection{The scalar-vector interaction model - SVI}
\label{svi}

The scalar-vector interaction (SVI) model has recently been 
developed by the author \cite{Sharma.08} with a view to remove nonlinearities 
of the mesonic fields in the RMF theory. Self-interactions of $\sigma$ field 
have been cited as a source of instability in the nuclear matter 
at higher densities \cite{Sulaksono.07}. 
It has now been demonstrated \cite{Sharma.08} that 
by a suitable combination of couplings of $\sigma$ and $\omega$ fields, 
the self-interactions of the mesonic fields can be dispensed with. 
It is expected that the introduction of SVI would be consistent with 
a linear realization of the chiral symmetry in a suitable scheme 
\cite{Fomenko.93}. 

The SVI model consists of the {\it meson-meson interactions}
between $\sigma$ and $\omega$ mesons of the form
\begin{equation}
\label{svieq}
U_{mm} = \frac{1}{2}g_4 \sigma\omega^2 + 
\frac{1}{2}g_5 \sigma^2\omega^2,
\end{equation}
where $g_4$ and $g_5$ are the respective coupling constants for
interactions between $\sigma$ and $\omega$ mesons. The effective 
Lagrangian {\it without self-interactions of the bosonic fields} thus becomes
\begin{equation}
\label{efflan}
{\cal L}_{eff} = {\cal L}_0 + U_{mm}.
\end{equation}
The corresponding Klein-Gordon equations (Eq. \ref{KG1}) have the
effective meson masses as given by:
\begin{equation}
\label{effm3}
\begin{array}{rl}
m_{\sigma}^{*2} = & m_{\sigma}^2  - g_4{\omega_0}^2/(2\sigma) 
               -  g_5{\omega_0}^2 \\
m_{\omega}^{*2} = & m_{\omega}^2 + g_4 \sigma  + g_5 \sigma^2\\
m_\rho^*        = & m_\rho.

\end{array}
\end{equation}
These equations exhibit an implicit density dependence
of the $\sigma$ and $\omega$ meson masses. Both the $\sigma$ and $\omega$ meson
masses provide a density dependence that is different from that for 
the SIGO model (Eq. \ref{effm2}). Thus, the density dependence of
$\sigma$ and $\omega$ meson masses, as represented by Eqs. (\ref{effm1}),
(\ref{effm2}) and (\ref{effm3}) for the three Lagrangian models NL$\sigma$,
SIGO and SVI, respectively, are different from one another. A comprehensive
discussion of the differences in the density dependences in various
Lagrangian models and their influence on nuclear properties shall be
presented elsewhere \cite{SS.08}. It shall be instructive to see as to how
these density dependences in the high-density regime would compare with 
the predictions of the Brown-Rho scaling \cite{BR.91}.

\noindent
\begin{table}[h*]
\begin{center}
\caption{The nuclear matter properties of the scalar-vector interaction (SVI) 
sets.}
\bigskip
\begin{tabular}{l l c c c c c }
\hline
& Sets     & $K_\infty$ (MeV) & $E/A$ (MeV) & $m^*$ & $\rho_0$
(fm$^{-3}$) & $J$ (MeV) \\    
\hline 
& SVI-a             & 243.5 & -16.30 & 0.615   & 0.150  & 37.4 \\  
& SVI-b             & 253.1 & -16.24 & 0.612   & 0.150  & 36.6 \\
& SVI-1             & 263.9 & -16.30 & 0.616   & 0.149  & 37.6 \\
& SVI-2             & 271.5 & -16.31 & 0.621   & 0.149  & 37.0 \\ 
& SVI-c             & 288.1 & -16.25 & 0.617   & 0.148  & 36.4 \\
\hline
\end{tabular}
\end{center}
\vspace{0.5cm}
\end{table}

The parameters sets SVI-1 ($K_\infty = 264$ MeV) and 
SVI-2 ($K_\infty = 272$  MeV) were obtained in Ref. \cite{Sharma.08}.
These sets have been shown to provide an excellent description of the 
ground-state energies of nuclei over a large range of the periodic table. 
This includes nuclei along the stability line as well as far away from it. 
Description of binding energies and charge radii of nuclei that is
achieved with the SVI sets is significantly better than that with the
NL$\sigma$ set NL3. Having established the basis of the SVI model, 
we have constructed a few more parameter sets having $K_\infty$ in the 
range $\sim 240-290$ MeV for a systematic analysis of the 
breathing-mode GMR. The nuclear matter properties of the parameter 
sets SVI-a, SVI-b, and SVI-c thus constructed in addition to
those of SVI-1 and SVI-2 are listed in Table 3. It is worth mentioning 
that the auxiliary sets SVI-a, SVI-b, SVI-c are also able to provide 
a reasonably good description of the ground-state properties of nuclei.

\section{The generator coordinate method and the breathing-mode GMR}
\label{GCM}

The generator coordinate method (GCM) is a powerful tool to
study ground and excited states in atoms and nuclei. The GCM has been
applied to study effects of correlations on the ground-state
properties of nuclei using the density-dependent Skyrme approach 
\cite{Bonche.91}. In the present work, we employ the GCM in the RMF theory
to explore the excited state of the isoscalar GMR in atomic nuclei. 

The GCM is based upon a trial $A$-particle wavefunction $\Psi_{GCM}$
in the form of a linear combination of
\begin{equation}
\Psi_{\rm GCM} ({\rm\bf r_1, r_2, ..., r_A}) = \int{{\cal{F}}(q) 
               \Phi({\rm\bf r_1, r_2, ..., r_A;}q)dq},
\end{equation}
where the generator function $\Phi(q) \equiv \Phi({\rm\bf r_1, r_2, ..., r_A};q)$
is a Slater determinant $\Phi(q)$ constituted from single-particle spinors
$\psi_i({\rm\bf r},q), (i = 1, 2, ..., A)$ as a function of the generator 
coordinate $q$. The ``weight'' or  the ``generator function'' ${\cal F}(q)$ is 
determined from variation of the total energy of the system 
\begin{equation}
E[{\cal F}] = \frac{\langle\Psi_{\rm GCM} \vert \hat{H} \vert \Psi_{\rm
    GCM}\rangle} {\langle\Psi_{\rm GCM} \vert \Psi_{\rm GCM}\rangle}
\end{equation}
with respect to ${\cal F}(q)$. This leads to the Hill-Wheeler equation for 
the weight function:
\begin{equation}
\label{HW}
\int[{\cal H}(r;q, q') - E{\cal N}(r;q, q')]{\cal F}(q') dq' = 0,
\end{equation}
where 
\begin{equation}
\label{ekern}
{\cal H}(r;q, q') = \langle\Phi(q) \vert \hat{H} \vert \Phi (q')\rangle
\end{equation}
is the overlap energy-density  kernel of the Hamiltonian $\hat{H}$ 
associated with the RMF Lagrangian, and
\begin{equation}
\label{olap}
{\cal N}(r;q, q') = \langle\Phi(q) \vert \Phi (q')\rangle
\end{equation}
is the overlap norm kernel. The overlap energy-density kernel takes
the form given in Eq. (\ref{hamilton}):
\begin{eqnarray}
\label{hamdens}
{\cal H}_{RMF}(r;q,q') &=& \tau(r;q,q') + M\rho_s(r;q,q') \nonumber\\
        && + \frac{1}{2} g_\sigma\rho_s(r;q,q')\sigma(r;q,q') 
         +\frac{1}{2}g_\omega\rho_v(r;q,q')\omega^0(r;q,q') \nonumber \\
&& +\frac{1}{2}g_\rho\rho_3(r;q,q')\rho^0(r;q,q') +
            \frac{1}{2}e\rho_p(r;q,q')A^0(r)\nonumber\\
&& +\frac{1}{2}((\nabla\sigma(r;q,q'))^2 + m_\sigma^2\sigma^2(r;q,q')\nonumber\\
&& -\frac{1}{2}((\nabla\omega^0(r;q,q'))^2 + m_\omega^2(\omega^0(r;q,q'))^2\nonumber\\
&& -\frac{1}{2}((\nabla\rho^0(r;q,q'))^2 + m_\rho^2(\rho^0(r;q,q'))^2\nonumber\\
&& -\frac{1}{2}((\nabla A^0(r;q,q'))^2 \nonumber\\
&& + ~{\rm other~terms}.
\end{eqnarray}
Here, the 'other terms' refer to the meson-meson interaction terms of various
Lagrangian models considered in Section \ref{RMF}. The kinetic energy density 
can be written in terms of the spinors $\{\psi_i({\bf\rm r};q)\}$ as
\begin{equation}
\tau(r;q,q') = \sum N^{-1}_{ji} \psi_i^\dagger({\bf r};q)
                  \{-i{\bf \alpha\nabla}\} \psi_j({\bf r};q')
\end{equation}
The source densities appearing in Eq. (\ref{hamdens}) can be written as 
\begin{eqnarray}
\rho_s(r;q,q')~~& =& ~~\sum N_{ji}^{-1}\bar{\psi_i}({\bf r};q)
\psi_j({\bf r};q'),\nonumber\\
\rho_v(r;q,q')~~& =& ~~\sum N_{ji}^{-1}\psi_i^\dagger({\bf r};q)
\psi_j({\bf r};q'),\nonumber\\
\rho_3(r;q,q')~~& =& ~~\sum N_{ji}^{-1}\psi_i^\dagger({\bf r};q)\tau_3
\psi_j({\bf r};q'),\nonumber\\
\rho_p(r;q,q')~~& =& ~~\sum N_{ji}^{-1}\psi_i^\dagger({\bf r};q)
\frac{(1-\tau_3)}{2}\psi_j({\bf r};q').
\end{eqnarray}
The overlap matrix elements $N_{ij}(q,q')$ are calculated using the spinors as
\begin{equation}
N_{ij}(q,q') = \int d^3r ~\psi_i^\dagger({\bf r};q) \psi_j({\bf r};q')
\end{equation}
The determinant of $N_{ij}(q,q')$ provides the overlap kernel (Eq. \ref{olap}):
\begin{equation}
\label{norm}
{\cal N}(q,q') = det\{N_{ij}(q,q')\}.
\end{equation}
On obtaining the energy and norm overlap kernels using Eqs. (\ref{ekern}) and
(\ref{norm}), the Hill-Wheeler equation (\ref{HW}) is solved. The solution 
provides the nuclear ground and the excited states. 

In order to obtain the isoscalar GMR energy of nuclei, we perform 
constrained RMF calculations by solving the Dirac equation
\begin{equation}
(\hat{h}_D - q\hat{Q})\psi_i(x) = \epsilon_i\psi_i(x)
\end{equation} 
with the constraint operator $\hat{Q} = r^2$. For the case of the
isoscalar GMR, the Lagrange multiplier $q$ is associated with the value
of the nuclear root-mean-square ($rms$) radius
\begin{equation}
R = \langle \hat{Q}\rangle = \left\{\frac{1}{A}\int r^2\rho_v(r;q)d^3r\right\}^{1/2}
\end{equation}
where $\rho_v(r;q)$ is the baryon density determined by the solution
$\{\psi_i({\bf r};q)\}$ for a given value of the generator coordinate $q$.

Constrained RMF calculations have been performed in coordinate space 
for a range of  the Lagrange multiplier $q$. Constructing the generator 
Slater determinant $\Phi(q)$, the integral kernels, Eq. (\ref{ekern}) 
and Eq. (\ref{olap}) are obtained. On solving the Hill-Wheeler equation, 
the first excited state gives the energy of the isoscalar GMR. 
Details of the GCM method applied to the RMF theory have been 
provided in Ref. \cite{Sto.94}.

\section{Results and discussion}
\label{results}

As noted earlier in Section \ref{RMF}, the Lagrangian sets SIG-OM in the 
SIGO model \cite{Haidari.08} and SVI-1 and SVI-2 in the 
SVI model \cite{Sharma.08} developed recently have been 
shown to bring about a significant improvement 
in the ground-state properties of nuclei as compared to 
the parameter set NL3 of the NL$\sigma$ model. 
With the advent of the alternative Lagrangian  models SIGO 
and SVI, we have now explored the response of various  
Lagrangian models to the breathing-mode isoscalar GMR. 

As alluded to in Ref. \cite{Haidari.08}, the set SIG-OM 
yields GMR energies which are larger than those of NL3, though the 
incompressibility $K_\infty$ of SIG-OM is smaller 
than that of NL3. It seems paradoxical, for according to conventional 
wisdom, a force with a higher value of $K_\infty$ should yield a 
higher value of the GMR energy and vice versa. This strange feature
has prompted us to investigate the GMR response within the framework of the 
RMF theory employing various Lagrangian models at hand. The isoscalar GMR 
energies have been calculated for a few key nuclei using the GCM approach.
\begin{figure}[h*]
\vspace {0.5 cm}
\hspace{2.0cm}
\resizebox{0.55\textwidth}{!}{%
   \includegraphics{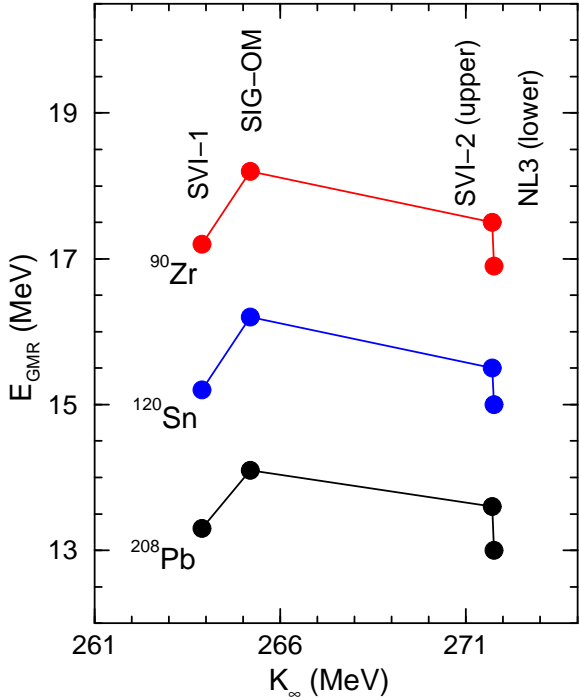}}
\vspace{0cm}       
\caption{The breathing-mode GMR energies for nuclei 
$^{208}$Pb, $^{120}$Sn and  $^{90}$Zr obtained with 
the generator coordinate method using the NL$\sigma$  
Lagrangian set NL3, the SVI Lagrangian sets SVI-1 and 
SVI-2 and the SIGO Lagrangian set SIG-OM. The sets NL3 and
SVI-2 have nearly the same value of $K_\infty$.}
\label{fig:1}       
\end{figure}

\subsection{The GMR energies: a paradoxical picture}

In order to visualize the response of various Lagrangian models,
we show  the GMR energies obtained by employing 
the GCM approach in Fig. \ref{fig:1}. The GMR energies for three key 
nuclei of $^{208}$Pb, $^{120}$Sn and $^{90}$Zr 
have been obtained using the parameter set NL3 of the NL$\sigma$ model, 
with the sets SVI-1 and SVI-2 of the SVI model \cite{Sharma.08}
and with the set SIG-OM of the SIGO model \cite{Haidari.08}. Contrary  
to the usual increase of the GMR energy with $K_\infty$, 
an unusual pattern displaying rather paradoxical behaviour is seen. 
The GMR energies obtained with SIG-OM are bigger
than those of NL3 as mentioned above. These are also bigger than those of
SVI-2. Moreover, SVI-2 values are bigger than those of NL3, though SVI-2 and
NL3 have nearly same value of $K_\infty$. This behaviour implies that
some finite-size contribution(s) to the incompressibility must be different  
in these Lagrangian models. Such a behaviour was noted in relativistic RPA
calculations in Ref. \cite{Ma.97}. It was shown that the force 
NL-SH of NL$\sigma$ model and TM1 with quartic $\omega$ coupling 
\cite{Suga.94}, both having very different values of $K_\infty$, 
were shown to provide comparable values of the GMR energy 
for $^{208}$Pb. In order to throw light upon the seemingly 
paradoxical behaviour of the GMR energies and with a view to gauge the 
response of the finite-size effects to the nuclear incompressibility 
of finite nuclei, we have first analyzed the GMR energies within 
each Lagrangian model separately.

\begin{figure}[h*]
\vspace {0.5 cm}
\hspace{0.5cm}
\resizebox{0.80\textwidth}{!}{%
   \includegraphics{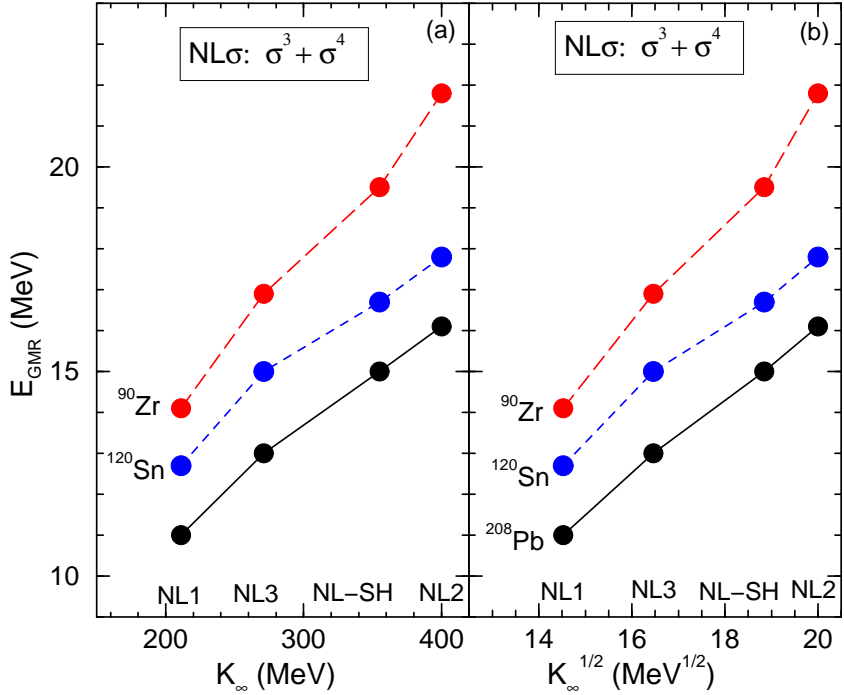}}
\vspace{0cm}       
\caption{The breathing-mode isoscalar GMR energies for $^{208}$Pb, 
$^{120}$Sn and
$^{90}$Zr obtained with the standard parameter sets NL1, NL3, NL-SH 
and NL2 of the NL$\sigma$ model in an increasing sequence of $K_\infty$.
The energies $E_{GMR}$ are displayed as a function of (a) $K_\infty$ and
(b) $\sqrt{K_\infty}$.}
\label{fig:2}       
\end{figure}
\subsection{GMR energies with the NL$\sigma$ model}
In the NL$\sigma$ model, we have employed the standard parameter sets
NL1, NL3, NL-SH and NL2, which are in an increasing sequence of 
$K_\infty$. These span a large range of $K_\infty \sim$ 210-400 MeV.
The isoscalar (T=0) GMR energies calculated for nuclei $^{208}$Pb, 
$^{120}$Sn and $^{90}$Zr using these parameter sets are shown
in Fig. \ref{fig:2}. The GMR energies are shown directly as a
function of $K_\infty$ in Fig. \ref{fig:2}(a). Due to obvious
dependence of $E_{GMR}$ on $K_A$, the curves exhibit a 
quadratic like dependence on $K_\infty$. Only for $^{90}$Zr does
the energy seems to fall slightly out of the trend especially for
NL2. This may be due to an extremely large value of $K_\infty \sim 400$ 
MeV for NL2, whereby some anharmonicity may creep in for $^{90}$Zr, 
which is relatively light as compared to $^{208}$Pb. 
It is interesting to note that the curve for each 
nucleus and especially that for the heavy nucleus $^{208}$Pb shows 
a monotonic dependence on $K_\infty$, though each of the parameter 
sets was constructed under different circumstances. Some nuclear matter 
properties of these sets such as the asymmetry energy are known to be 
rather different especially that of NL1 and NL2 with a value of 
$J\sim$44 MeV. 

The GMR energies are shown as a function of $\sqrt{K_\infty}$ in
Fig. \ref{fig:2}(b) in order to eliminate the apparent quadratic
dependence. The curve for $^{208}$Pb shows a slight improvement
towards a linear behaviour. For $^{120}$Sn and $^{90}$Zr a 
similar feature may not be visible due to interplay of 
stronger finite-size effects such as the surface which 
becomes important for medium heavy and lighter nuclei. 
\subsection{The GMR energies with the SVI model}

\begin{figure}[h*]
\vspace {0.5 cm}
\hspace{2.5cm}
\resizebox{0.60\textwidth}{!}{%
   \includegraphics{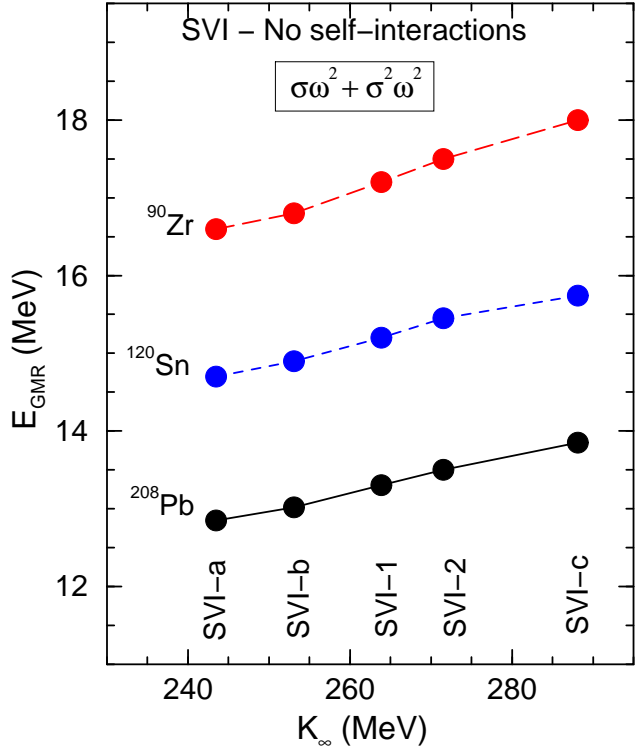}}
\vspace{0cm}       
\caption{The breathing-mode isoscalar GMR energies for $^{208}$Pb, 
$^{120}$Sn and $^{90}$Zr obtained with the parameter 
sets SVI-a, SVI-b, SVI-1, SVI-2, and SVI-c of the scalar-vector 
interaction model SVI of Ref. \cite{Sharma.08}.}
\label{fig:3}       
\end{figure}

We have calculated the breathing-mode GMR energies within the 
scalar-vector Lagrangian model  SVI using the GCM with the 
parameter sets SVI-a, SVI-b, SVI-1, SVI-2 and SVI-c, which are 
in an increasing order of $K_\infty$.  These sets encompass a region of  
$K_\infty \sim 240-290$ MeV. The results are shown in Fig. \ref{fig:3}.
The curves show a monotonous behaviour with $K_\infty$. All the parameter
sets of SVI fall in an orderly pattern and show a dependence on
$K_\infty$ similar to that seen in Fig. \ref{fig:2}. However, as compared to 
the values in  Fig. \ref{fig:2}, the GMR energies with SVI are 
higher than the corresponding values with the NL$\sigma$ model. 
\subsection{The GMR energies with the SIGO model}

\begin{figure}[h*]
\vspace {0.5 cm}
\hspace{2.5cm}
\resizebox{0.60\textwidth}{!}{%
   \includegraphics{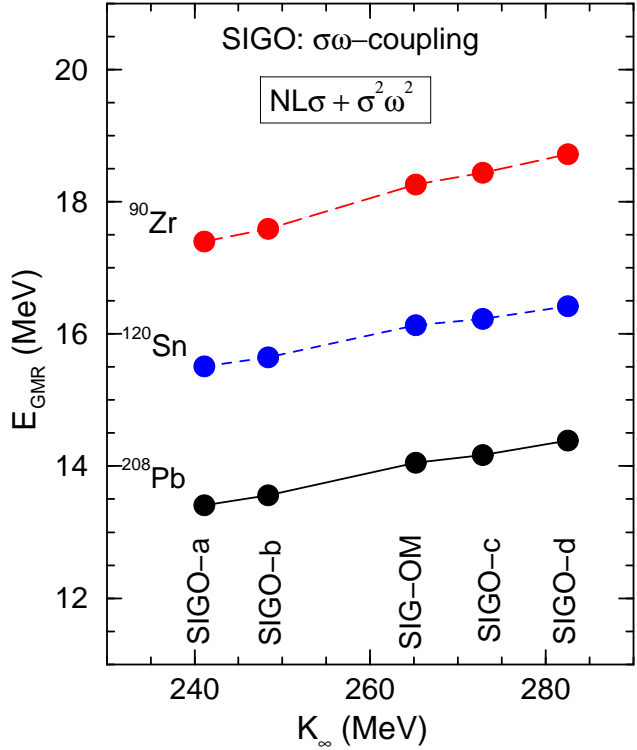}}
\vspace{0cm}       
\caption{The GMR energies for $^{208}$Pb, $^{120}$Sn and $^{90}$Zr 
obtained with the parameter sets SIGO-a, SIGO-b, SIG-OM, SIGO-c and 
SIGO-d of the $\sigma$-$\omega$ model SIGO of Ref. \cite{Haidari.08}}
\label{fig:4}       
\end{figure}

We have then investigated the response of the $\sigma$-$\omega$ model
SIGO to the breathing-mode GMR. GCM calculations have been 
performed with the parameter sets 
SIGO-a, SIGO-b, SIG-OM, SIGO-c and SIGO-d. The results are shown in
Fig. \ref{fig:4}. All the SIGO sets display an orderly pattern
such as that seen in Figs. \ref{fig:2} and \ref{fig:3} with the NL$\sigma$
and SVI models, respectively. Comparatively, the SIGO 
values are higher than the corresponding SVI values.

\subsection{A comparative picture}

\begin{figure}[h*]
\vspace {0.5 cm}
\hspace{2.5cm}
\resizebox{0.55\textwidth}{!}{%
   \includegraphics{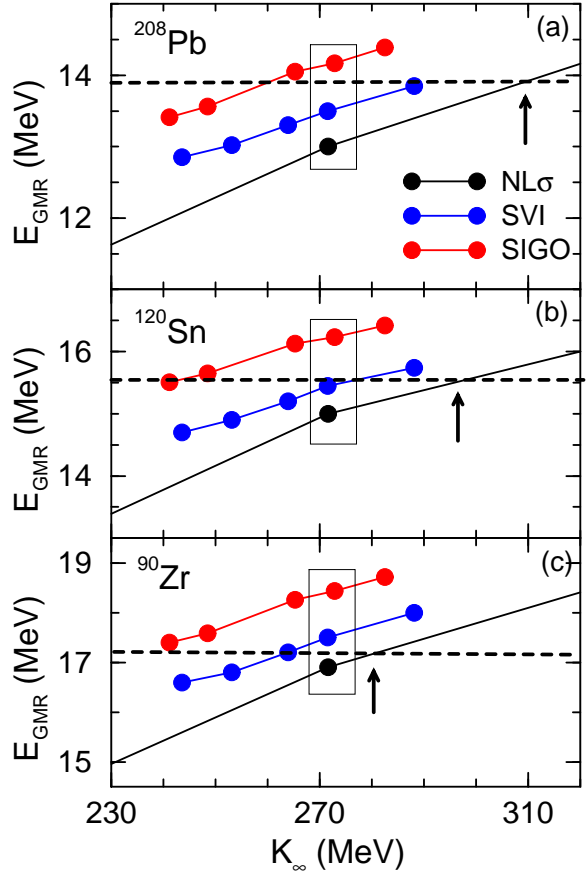}}
\vspace{0cm}       
\caption{The GMR energies for $^{208}$Pb, $^{120}$Sn and
$^{90}$Zr for the three Lagrangian models in the physically 
acceptable region of $K_\infty$ compared with the experimental 
data. The arrows show intersection of the experimental data 
(dashed lines) with the theoretical curves obtained with the 
NL$\sigma$ model. The results with sets having nearly the same
value of $K_\infty \sim 272$ MeV in the three Lagrangian models 
are enclosed by rectangular boxes.}
\label{fig:5}       
\end{figure}

The results obtained with the three Lagrangian models
are compared in Fig. \ref{fig:5}. Focusing on the region of physically 
acceptable values of $K_\infty$, the figure shows the GMR energies for 
the three Lagrangian models within the range of $K_\infty \sim 230-320$ MeV.
For a given value of $K_\infty$, the model NL$\sigma$ gives the 
lowest values, whereas the model SIGO delivers the largest values 
for the breathing-mode GMR energy. Thus, the models NL$\sigma$, SVI and 
SIGO, respectively, produce GMR energies in an increasing order for any 
given nucleus. The results in this figure demonstrate that the GMR energies 
depend strongly upon the Lagrangian model employed and that each model 
predicts a different value of the GMR energy for a given $K_\infty$. 
This implies that the GMR energy is not a simple function 
of only the incompressibility $K_\infty$ of the infinite nuclear matter.
Different values of the GMR energy for a given value of 
$K_\infty$ indicate different contribution of 
finite-size effect(s) to the
incompressibility $K_A$ of a nucleus. We discuss below the 
relative magnitudes of various contributions to $K_A$ and attempt to
discern the factors responsible for multiplicity of the GMR energies for a
given value of $K_\infty$ and thus the reason for producing the 
paradoxical behaviour seen in Fig. \ref{fig:1}. 

The multiplicity of the GMR energies for a given $K_\infty$
in the RMF theory is in contrast to a single value that is 
obtained usually within the density-dependent Skyrme 
theory \cite{Blaizot.95}. This can be attributed to a difference 
in the density dependence of the nuclear interaction amongst 
the various Lagrangian models. In comparison, 
the density dependence within the non-relativistic Skyrme approach 
is provided by the standard Skyrme density functional used
universally - which is fixed at the outset. Should an alternative 
form  of the density functional in the Skyrme approach arise, a situation 
akin to the RMF theory would emerge. Microscopic differences amongst
various Lagrangian models, which lead to a phenomenal difference in the 
GMR energies, are being investigated separately \cite{SS.08}. 

\subsection{Theoretical extraction of $K_\infty$ - a conundrum}

Given the emerging situation in the RMF theory, an intersection of the 
experimental values (shown by the dashed horizontal lines in 
Fig. \ref{fig:5}) with the theoretical curves would yield a 
different value of $K_\infty$ depending upon the Lagrangian model used. 
The experimental values shown by the dashed lines 
cross the curves at different locations in $K_\infty$. 
Staying with the pivotal case of $^{208}$Pb in 
Fig. \ref{fig:5}(a), each Lagrangian model yields $K_\infty$ value
which is different from the others. $K_\infty$ inferred from 
the datum of $^{208}$Pb with the model NL$\sigma$ is $\sim 310$ MeV,
whereas a value of $\sim 290$ MeV and $\sim 260$ MeV would be 
determined with the models SVI and SIGO, respectively.

Using the datum on $^{120}$Sn in Fig. \ref{fig:5}(b), the value of
$K_\infty$ inferred is $\sim300$ MeV,  $\sim275$ MeV and
$\sim245$ MeV with the NL$\sigma$, SVI and SIGO models, respectively.
Similarly, using the datum on $^{90}$Zr in Fig. \ref{fig:5}(c),
one would conclude a value of $K_\infty$ as $\sim280$ MeV,  
$\sim265$ MeV and $\sim240$ MeV, respectively. The shift in the value
of $K_\infty$ inferred from a given model using the experimental data 
on $^{208}$Pb, $^{120}$Sn and $^{90}$Zr is illustrated by the arrows 
for the case of the NL$\sigma$ model. A similar shift is evident in
the figure (not shown by arrows) for the models SVI and SIGO as well.

Staying with the interpolation method, it is equally illustrative to 
conclude that any of the Lagrangian models is not able to match the
experimental data on all the three nuclei simultaneously with a given 
value of $K_\infty$. These observations bring into question the hitherto 
used theoretical approach based upon interpolation amongst forces. 
Thus, the approach of determining $K_\infty$ \'{a} la Blaizot 
\cite{Blaizot.80} becomes unsustainable in the RMF theory. It is therefore 
essential to 'calibrate' the density dependence of the nuclear 
interaction in the RMF theory before such an approach could be
applicable. Arguably, the form of the density dependence opens up 
another dimension in the landscape of the breathing-mode GMR  and 
the incompressibility of nuclear matter.

The shift of the arrows from the right towards the left 
in going from $^{208}$Pb to $^{90}$Zr (i.e., a decrease 
in the value of $K_\infty$ so inferred) needs analysis. 
It is understood that the interior (or bulk) of the 
nucleus $^{208}$Pb has a density close to that of the saturation
density of nuclear matter and hence is synonymous with the nuclear
matter for practical purposes. It has a nuclear surface that 
is smaller as compared to that in $^{120}$Sn and $^{90}$Zr. 
Thus, in going from $^{208}$Pb
to $^{90}$Zr, the region of the nuclear surface increases. The 
effect of compression of a correspondingly large nuclear surface
needs to be accounted for. The continuous shift in the arrows (and hence
$K_\infty$ value inferred) from $^{208}$Pb to $^{90}$Zr points towards the 
increased role played by an increasingly larger surface region. 

\subsection{Disentangling the finite-size effects and the surface 
incompressibility}

A shift in the value of $K_\infty$ extracted from the experimental
GMR energies of $^{208}$Pb, $^{120}$Sn and $^{90}$Zr, as shown
in Fig. \ref{fig:5}, renders the approach of interpolation amongst 
forces as futile. It is, therefore, pertinent to revisit the expansion 
of $K_A$ to dissect various components. As noted earlier, the use of 
Eq. (\ref{KA})  does not allow an unambiguous determination of 
$K_\infty$ primarily due to the correlation of the Coulomb 
term to $K_\infty$ superimposed on to a much smaller magnitude 
of the former. Here, we adopt a slightly different approach.
Instead of allowing the aforesaid correlation to jeopardise the 
extraction of $K_\infty$, we let the coefficient $K_{Coul}$ 
fit as an independent variable.

For a comparative analysis of the GMR energies, we have selected the 
parameters sets having the same value of $K_\infty$  in the three 
different Lagrangian models.  The sets with an incompressibility 
close to $K_\infty \sim 272$ MeV are enclosed by the rectangular 
boxes in Fig. \ref{fig:5}. This corresponds to the sets NL3 
($K_\infty = 271.8$ MeV), SVI-2 ($K_\infty = 271.4$ MeV) and 
SIGO-c ($K_\infty = 272.8$ MeV) in the Lagrangian models 
NL$\sigma$, SVI and SIGO, respectively. 

\noindent
\begin{table}[h*]
\begin{center}
\caption{The GMR energies for $^{208}$Pb, $^{90}$Zr, $^{48}$Ca, and $^{40}$Ca
obtained with the parameter sets NL3, SVI-2 and SIGO-c having the same
value of $K_\infty \sim 272$ MeV in the three Lagrangian models.}
\bigskip
\begin{tabular}{c c c c c c }
\hline
& Nucleus    & NL3  & SVI-2 & SIGO-c \\    
\hline 
& $^{208}$Pb & 13.0 & 13.5  & 14.2  \\
& $^{90}$Zr  & 16.9 & 17.5  & 18.4  \\ 
& $^{48}$Ca  & 18.9 & 19.7  & 20.5 \\
& $^{40}$Ca  & 19.6 & 20.3  & 21.1 \\
\hline
\end{tabular}
\end{center}
\vspace{0.5cm}
\end{table}

We have selected the key nuclei of $^{208}$Pb, $^{90}$Zr, $^{48}$Ca, and 
$^{40}$Ca spanning a broad range of masses for our analysis. $^{48}$Ca and 
$^{40}$Ca are included with a view to augment the range of variation 
with $A$ and to reinforce the asymmetry component, and thus to facilitate 
a reasonable dissection of the components in the analysis. As observed 
in Refs. \cite{Li.07,Pieka.07,Garg.07}, Sn isotopes present an anomalous
behaviour and are at present problematic theoretically. 
Sn nuclei being open-shell exhibit a significant BCS neutron
superfluidity in the ground state thus affecting the Fermi 
surface significantly. These may require further theoretical 
considerations than employed so far. With the problem of Sn nuclei 
not yet understood, we have not included $^{120}$Sn isotope in 
the dissection of finite-size contributions.

The GMR energies for $^{208}$Pb, $^{90}$Zr, $^{48}$Ca, and $^{40}$Ca  
obtained with the GCM using the parameter sets NL3, SVI-2 and 
SIGO-c of the three Lagrangian models are given in Table 4. 
As seen earlier, NL3, SVI-2 and SIGO-c, respectively, produce GMR 
energies in an increasing order for all the nuclei considered.
The energies with NL3 are the lowest, whereas SIGO-c provides the 
largest values amongst the three Lagrangian models. A preliminary
consideration suggests that $K_\infty$ being the same for these 
parameter sets, the finite-size contributions to Eq. (\ref{KA}) 
must be largest with NL3 and smallest with SIGO-c in order 
to have the noted difference in the GMR energies. 

In order to discern which finite-size effect(s) play a
crucial role in influencing the GMR energies such as given 
in Table 4 and portrayed in Figs. \ref{fig:5},
we have resorted to the use of Eq. (\ref{KA}). The objective is to
see whether a fit to Eq. (\ref{KA}) is possible by relinquishing the
aforesaid correlation of $K_{Coul}$ to $K_\infty$. Using the GMR 
energies given in Table 4 for the parameters sets of the three 
Lagrangian models, we have performed a $\chi^2$-minimization procedure 
to fit Eq. (\ref{KA}) by treating $K_{surf}$, $K_{asym}$, $K_{Coul}$ 
and $K_{curv}$ as independent variables. The corresponding $K_A$ is 
evaluated using Eq. (\ref{E0}).

\noindent
\begin{table}[h*]
\begin{center}
\caption{The coefficients of expansion (in MeV) of $K_A$ for the parameter 
sets of the three Lagrangian models with the same value of $K_\infty \sim 272$
MeV. The last column shows the ratio $K_{surf}/K_\infty$ obtained.}
\bigskip
\begin{tabular}{l l c c c c c c}
\hline
& Model~ & ~~$K_\infty$~~  & ~~$K_{surf}$~~ & ~~$K_{asym}$~~ &
~~$K_{Coul}$~~ & ~~$K_{curv}$~ & $K_{surf}/K_\infty$ \\    
\hline 
& NL3       & 271.8 & $-537 \pm 15$ & $-389 \pm 12$ & $-6.9 \pm 0.2$  
  & $117 \pm 43 $ & $-$1.98 \\  
& SVI-2     & 271.4 & $-455 \pm 15$ & $-345 \pm 12$ & $-6.8 \pm 0.2$  
  & $-52 \pm 45$ & $-$1.67   \\
& SIGO-c    & 272.8 & $-273 \pm 17$ & $-395 \pm 13$ & $-7.1 \pm 0.3$ 
  & $-622 \pm 50$  & $-$1.00  \\
\hline
\end{tabular}
\end{center}
\vspace{0.5cm}
\end{table}

For the purpose of this work, i.e. to investigate the importance of 
finite-size effects, we fix the value of $K_\infty$ to the theoretical 
one and allow a fit of the GMR energies only to four free
parameters, viz., $K_{surf}$, $K_{asym}$, $K_{Coul}$ and 
$K_{curv}$. It is noted that by doing so it is possible to fit 
Eq. (\ref{KA}) to the GMR energies well. The results of the 
fits of the GMR energies for the three Lagrangian models 
are given in Table 5. The resulting coefficient $K_{Coul}$ turns out 
be similar in value from NL3 to SIGO-c. It is not expected that the 
Coulomb term would be very different from one Lagrangian model to 
the other. These values are very close to those which have  been 
obtained using other approaches such as the density-dependent 
Skyrme forces \cite{Sagawa.07}. 

\begin{figure}[h*]
\vspace {0.5 cm}
\hspace{2.0cm}
\resizebox{0.60\textwidth}{!}{%
   \includegraphics{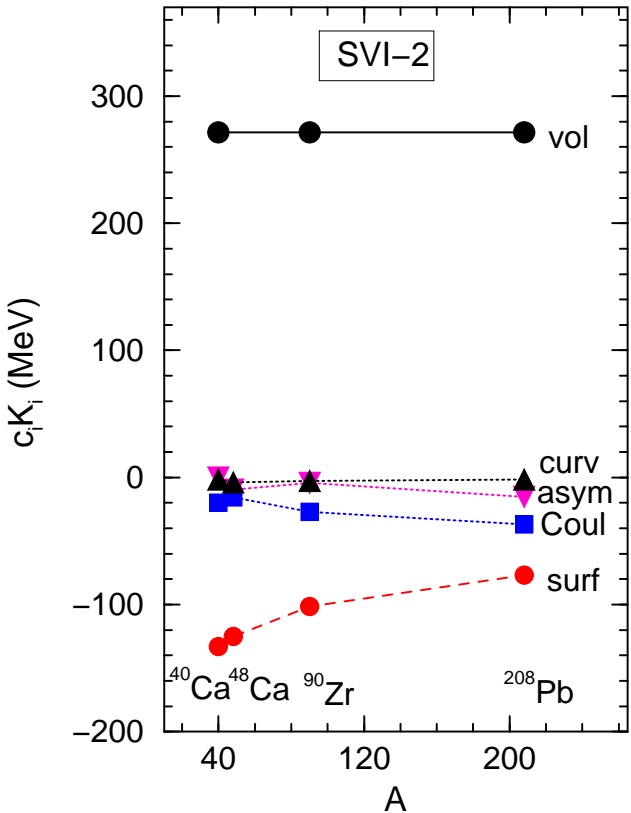}}
\vspace{0cm}       
    \caption{Contributions of various terms in the expansion 
   $K_A = \Sigma c_iK_i$ using the coefficients $K_i$ from Table 5 
   for the parameter set SVI-2.}
\label{fig:6}       
\end{figure}

It can be seen from Table 5 that by fixing $K_\infty$ to the desired
value, the surface term can be determined with a considerable
precision. The striking feature that emerges from the minimizations 
is the significant difference in the value of the surface 
incompressibility. Here $K_{surf}$ is obtained as $-537$ MeV for NL3,  
$-455$ MeV for SVI-2 and $-273$ MeV for SIGO-c. The set NL3 provides 
the largest value of $K_{surf}$, whereas SIGO-c yields the lowest value.  
Inevitably, a large difference in the value of 
$K_{surf}$ for the three Lagrangian models is responsible 
for the differences in the GMR energies as observed in Figs. \ref{fig:5} 
and \ref{fig:6}. This is also reflected by the significant differences 
in the ratio $K_{surf}/K_\infty$ shown in the last column. It varies 
from $-1.98$ for NL3, $-1.67$ for SVI-2 to $-1.00$ for SIGO-c. 
The ratio with NL3 and SVI-2 is substantially different from the 
ratio of $\sim -1$ that is obtained from calculations of semi-infinite 
nuclear matter using the scaling model \cite{Patra.02}. On the other
hand, a ratio of $\sim -1$ is attained with SIGO-c. 

In comparison, the asymmetry coefficient $K_{asym}$  shows only a 
limited variation from $\sim -345$ MeV to $\sim -395$ MeV (Table 5). 
These values are not far from the value of $-550 \pm 100$ MeV 
obtained in a recent fit of the experimental GMR energies 
of Sn isotopes \cite{Li.07}. A  value of $-500 \pm 50$ MeV
has been obtained in a recent analysis of the breathing-mode
GMR with the density-dependent Skyrme forces \cite{Sagawa.07}.

For the sake of illustration of the relative magnitudes of various 
contributions to $K_A$ in Eq. (\ref{KA}), we show the size of various 
terms using the coefficients obtained for the parameter set SVI-2 in 
Fig. \ref{fig:6}. Here we have chosen the set SVI-2 for the sake of
illustration. The picture with the other parameters sets is similar 
to it with major differences being in the surface contribution. 
The curvature term would show differences as maifest in Table 5
especially for SIGO-c. The variation of the Coulomb and the asymmetry 
terms with the mass number $A$  is relatively small. Additionally, 
the Coulomb and the asymmetry contributions are an order 
of magnitude smaller than the surface term. 

The surface term provides the largest contribution to $K_A$ amongst 
the finite-size effects. It also shows a significant variation with $A$.
Consequently, the surface incompressibility has the ability to modulate 
the GMR energies of medium mass and lighter nuclei in a significant measure.

\begin{figure}[h*]
\vspace {0.5 cm}
\hspace{2.0cm}
\resizebox{0.60\textwidth}{!}{%
   \includegraphics{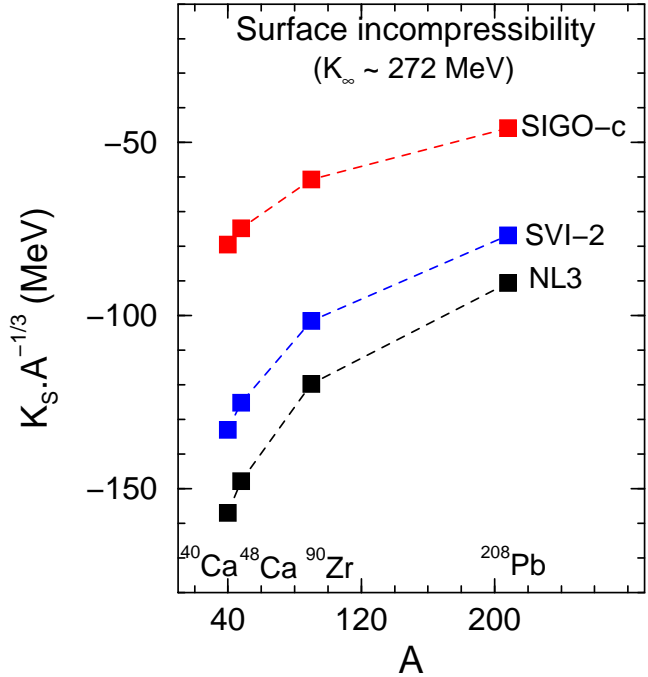}}
\vspace{0cm}       
\caption{The surface contribution to $K_A$ for the parameter 
sets NL3, SVI-2 and SIGO-c in the three Lagrangian models
having the same value of $K_\infty \sim 272$ MeV.}
\label{fig:7}       
\end{figure}

The relative magnitudes of the surface term for the parameter sets
of the three Lagrangian models having the same value of 
$K_\infty \sim 272$ MeV are compared in Fig. \ref{fig:7}. The surface
term exhibits a large variation with mass $A$. Significant 
differences in the surface term amongst the various Lagrangian
models are seen readily in the figure. Inevitably, these differences
give rise to the differences in the breathing-mode energies obtained 
with various Lagrangian models. 

Given the relatively smaller magnitude of the Coulomb
and the asymmetry terms as seen in Fig. \ref{fig:6}, magnitudes of 
these terms would not show any significant difference from one 
Lagrangian model to the others. In conclusion, it is the surface 
term that plays a decisive role in influencing the GMR energies of nuclei 
amongst various Lagrangian models significantly. Whilst the curvature
term is not very small especially with SIG-c, it may also be influencing
the GMR energies to an extent.

\subsection{Density dependence of the meson masses}

The three Lagrangian models employed in this work are successful
in reproducing the ground-state properties of nuclei. Whilst
most of the nuclear matter properties e.g., saturation density,
saturation binding energy, effective mass and asymmetry energy are
not very different from one model to the other, the density dependence 
of the $\sigma$ and $\omega$ meson masses does show different forms in
various models as seen in Section \ref{RMF}. This difference in the density
dependence especially in the sub-saturation densities would influence 
the surface properties. Consequently, the 
form of the density dependence of meson masses affects 
the GMR response implicitly.

The density dependence of meson masses and by implication that of 
the nuclear interaction in the RMF theory remains an open and 
challenging problem. 
The introduction of the nonlinear $\sigma^3 + \sigma^4$ terms 
in the RMF Lagrangian has led to successful results on
ground-state properties of nuclei. It has, however, a major 
drawback in that the resulting EOS of the NL$\sigma$
model is ubiquitously stiff. The standard NL$\sigma$ sets produce
a maximum mass of neutron stars in the vicinity of 2.7-3.0 solar 
masses. This is well above the spectrum of the observed neutron
star masses. Thus, the NL$\sigma$ model is not deemed as suitable
for neutron star structure. In view of this, development of other 
Lagrangian models which could describe finite nuclei as well as nuclear
matter at higher densities becomes desirable.

The recent development of the SVI \cite{Sharma.08} and 
SIGO \cite{Haidari.08} models provide a much improved description 
of nuclear properties vis-a-vis the NL$\sigma$ model.
Notably, the SIGO model provides an excellent description of nuclei 
at the shell closures and also for nuclei in the extreme regions of
the periodic table \cite{Haidari.08}. It also delivers an EOS of 
nuclear matter which is softer than the NL$\sigma$ counterparts.
On the other hand, the Lagrangian model SVI is able to describe the 
ground-state properties of nuclei without self-interactions 
of the mesonic fields. It has been shown that introduction of terms
of scalar-vector type such as those in SVI would help alleviate the 
onset of instabilities in nuclear matter \cite{Sulaksono.07}. 
Arguably, the SVI model would  also be conducive to a linear 
realization of chiral symmetry. 

Notwithstanding the above, the RMF theory offers alternative 
possibilities of its density functional to be applicable to nuclei 
and nuclear matter. A natural question that arises is: 
what is the density dependence of the meson masses that is 
preferred by nature? At present we do not have an answer to 
this question. It is difficult to pinpoint as to which Lagrangian 
model is preferred by experimental data. A comprehensive investigation of 
the RMF theory with all its attendant problems, virtues and theoretical
constraints juxtaposed to a host of experimental data would be 
necessary in order to be able to unravel the form of the density 
dependence required by nature. 

\subsection{The established parameter sets and confrontation with 
GMR energies}

In view of the desirability of discerning the density dependence 
in the RMF theory, it is instructive to examine the results of
the GCM calculations on breathing-mode GMR using the standard 
parameter sets vis-a-vis the available experimental data. The
parameter sets which come into such a consideration are the ones
which are in the region of physically acceptable values of
$K_\infty$. This includes the sets NL1 and NL3 of the NL$\sigma$
model, sets SVI-1 and SVI-2 of the SVI model and SIG-OM of the SIGO
model. With the exception of NL1, all the other sets have been
shown to describe binding energies and charge radii of nuclei along
as well as far away from the stability line. 

The GMR energies of the key nuclei $^{208}$Pb, $^{120}$Sn and 
$^{90}$Zr obtained with the GCM approach using the above 
parameter sets are shown in Table 6. Experimental data on 
these nuclei are well established and are shown for a comparison. 
The parameter sets are listed in an ascending order of the GMR
energies. 

\noindent
\begin{table}[h*]
\begin{center}
\caption{The breathing mode GMR energies obtained with constrained 
GCM calculations using the established parameter sets. The experimental 
data \cite{Sharma.88,Youngblood.04} are shown for comparison. The
$K_\infty$ value (in MeV) for the sets is shown in parentheses.}
\bigskip
\begin{tabular}{c c c c c c c c}
\hline
  ~~Sets~~  &  NL1  & NL3    & SVI-1  & SVI-2  & SIG-OM &&  exp. \\    
            &  (211)& (272)  & (264)  & (272)  & (265)  &&   \\ 
\hline 
$^{208}$Pb & 11.0 & 13.0   & 13.3   & 13.5   & 14.1   && $13.96 \pm 0.28$ \\
$^{120}$Sn & 12.7 & 15.0   & 15.2   & 15.4   & 16.2   && $15.52 \pm 0.15$ \\
$^{90}$Zr  & 14.1 & 16.9   & 17.2   & 17.5   & 18.2   && $17.81 \pm 0.30$ \\
\hline
\end{tabular}
\end{center}
\end{table}

Without resorting to a disentanglement of $K_A$ into
its respective components, we compare the GMR energies due to 
various parameter sets directly with the experimental values.
The GMR energies with NL1 are $\sim$ 3-4 MeV smaller than
the experimental values. NL3, on the other hand, provides
GMR energies which are $\sim$ 1 MeV smaller than the experimental
ones for $^{208}$Pb and $^{90}$Zr. For $^{120}$Sn, the 
difference of the NL3 GMR energy with the experimental value
is $\sim$ 0.5 MeV. 

In comparison, the SVI-1 GMR energies are slightly bigger than
those of NL3 and hence are closer to the experimental values
than NL3. However, even with SVI-1, the difference with the experimental
data, especially for $^{208}$Pb and $^{90}$Zr is not insignificant.
On the other hand, the SVI-2 GMR energies for all the three nuclei
come much closer to the experimental bounds. As is the case with SVI-1,
the GMR energies for $^{208}$Pb and $^{90}$Zr with SVI-2 are 
on the lower side of the experimental values, whereas for $^{120}$Sn,
the SVI-2 value is in good agreement with the experimental datum.
Overall, the SVI-2 results show a reasonable good
agreement with the experimental data for the three nuclei.
The SIG-OM value, in comparison, comes closer to the 
experimental one for $^{208}$Pb. However, SIG-OM overestimates the 
experimental values for $^{120}$Sn by $\sim$ 0.7 MeV and slightly for
$^{90}$Zr. Thus, the behaviour of Sn isotope remains anomalous 
in the GCM approach, as observed in nonrelativistic and relativistic 
RPA approach \cite{Sagawa.07,Pieka.07}.

From the point of view of the ground-state binding energies
and charge radii of nuclei and a good description of data along 
the stability line and away from it, in conjunction with the 
breathing-mode GMR energies, the parameter set SVI-2 of the 
SVI model without self-interactions comes closest to the empirical
data. It has to be seen in further exploration of other aspects
of nuclear structure and properties as to how close to the data
the set SVI-2 will appear. It should be pointed out that this conclusion
drawn here is subject to the GCM treatment of the breathing mode
GMR. The constrained GCM approach delivers GMR energies slightly 
smaller than those obtained within the RPA approach. It is worth 
investigation as to how different the properties of the breathing-mode 
GMR would appear within the framework of relativistic 
RPA using various Lagrangian models.

\section{Summary and conclusions}
 
The breathing-mode GMR is investigated within the 
framework of the RMF theory using the generator coordinate 
method.  The GMR energy has been calculated for a few key 
nuclei of $^{208}$Pb, $^{120}$Sn, $^{90}$Zr, $^{48}$Ca and 
$^{40}$Ca covering a broad range of atomic mass. 
Using the Lagrangian set NL3 of the NL$\sigma$ model, 
sets SVI-1 and SVI-2 of the SVI model and set SIG-OM of 
the SIGO model, a paradoxical behaviour of the GMR energies 
has been found. Contrary to the received wisdom, parameter sets with a 
higher value of $K_\infty$ are observed to yield a lower value of 
the GMR energy as compared to a higher GMR energy attained with sets 
having a lower value of $K_\infty$. 

In order to resolve the apparent paradox of the GMR 
energies in the RMF theory, the GMR response has been investigated
using the three different Lagrangian models viz., the NL$\sigma$ 
model, the SVI model and the SIGO model. Employing the constrained
GCM approach, GMR energies have been calculated within each Lagrangian model 
using parameter sets which encompass a broad range 
of $K_\infty$. It is shown that each Lagrangian model exhibits a GMR 
response that is different from the others. Consequently, each 
Lagrangian model yields a distinctly different value of the GMR 
energy for a given value of $K_\infty$. The model 
NL$\sigma$ delivers the lowest GMR energy for a nucleus 
whereas the model SIGO produces the highest value amongst 
the three models for any given value of $K_\infty$. 

The GMR energy is shown to exhibit a  sensitivity to the Lagrangian model 
employed. This behaviour stems from the differences in the implicit
density dependence of the meson masses in various models, thus
altering the density dependence of the nuclear interaction in the RMF
theory from one Lagrangian model to the other. The sensitivity
to the density dependence opens up another dimension in the landscape 
of the GMR energy and incompressibility.  

The multiplicity of the GMR energy for a given $K_\infty$ in 
the RMF theory  renders the 'microscopic' method of 
extraction of $K_\infty$ using interpolation 
amongst forces as inapplicable. For this approach 
to work, however, it is necessary to 'calibrate' the density 
dependence of the meson masses and correspondingly the nuclear 
interaction in the RMF theory.

Using the liquid-drop type expansion of the finite nucleus incompressibility
$K_A$, it is shown that for a given value of $K_\infty$, each Lagrangian model 
yields a markedly different value of the surface incompressibility $K_{surf}$. 
The model NL$\sigma$ delivers the largest value of $K_{surf}$, whereas the 
model SIGO provides the smallest $K_{surf}$ amongst the three Lagrangian 
models considered. Consequently, a different value of the ratio 
$K_{surf}/K_\infty$ emerges in different models. Different $K_{surf}$ values 
arising in various Lagrangian models are shown to be largely responsible 
for a multitude of GMR energies for a given 
$K_\infty$. Thus, the response of the nuclear surface to compression 
is found to be dependent on the Lagrangian model employed. 
Additionally, the ratio $K_{surf}/K_\infty$ with the Lagrangian models
NL$\sigma$ and SVI emerges to be significantly larger than the ratio of 
$\sim -1$ that is usually obtained with the assumption of scaling in 
the breathing-mode GMR \cite{Patra.02}.

The GMR response in the non-relativistic Skyrme approach is simple in
contrast. The Skyrme approach commonly delivers a GMR energy which
depends  primarily on the value of $K_\infty$ \cite{Blaizot.95}.
This difference in the behaviour of the GMR energies between 
the RMF theory and the Skyrme approach can be attributed to the fact
that the form of the density dependence in the Skyrme density functional
is well prescribed at the outset in marked contrast to the various forms of
density dependence which have emerged in the RMF theory. 
In a forthcoming work, we shall shed light on the differences
amongst various Lagrangian models, which produce a varied response to
the GMR energy in the RMF theory shown in this work. Further investigations 
will also be necessary in order to be able to discern and formulate 
the appropriate density dependence in the RMF theory.

{\bf Acknowledgments} ~I thank Lev Savushkin and Roger Hilton for fruitful
discussions and for a careful reading of the manuscript. Thanks are due to 
the University of St. Petersburg Telecommunications, Russia and 
especially Lev Savushkin for hospitality for my stay in St. 
Petersburg during my sabbatical leave of absence, where part of this 
work was carried out. Useful exchanges with S.K. Patra are acknowledged 
with thanks.

\end{document}